\documentclass[11pt]{article}
\usepackage{url}
\usepackage{epsf}
\usepackage{epsfig}
\usepackage{graphicx}
\usepackage{amsfonts}
\usepackage{amsmath,amssymb}
\usepackage{latexsym}

\newcommand{\ABox}{
\raisebox{3pt}{\framebox[6pt]{\rule{6pt}{0pt}}}
}
\newenvironment{proof}{{\bf Proof:}}{\hfill\ABox}

\newtheorem{theorem}{{\bf Theorem}}

\newtheorem{lemma}[theorem]{Lemma}

\newcommand{\A}{\mathcal A}

\begin{document}

\title{An $O(n \log n)$-Time Algorithm for the Restricted Scaffold Assignment}
\author{
Justin Colannino 
\and Mirela Damian
\and Ferran Hurtado 
\and John Iacono
\and Henk Meijer 
\and Suneeta Ramaswami 
\and Godfried Toussaint}

\date{}

\maketitle

\begin{abstract}
The {\em assignment} problem takes as input two finite point sets
$S$ and $T$ and establishes a correspondence between points in $S$ 
and points in $T$, such that each point in $S$ maps to exactly 
one point in $T$, and each point in $T$ maps to at least one 
point in $S$. In this paper we show that this problem has an 
$O(n \log n)$-time solution, provided that the points in 
$S$ and $T$ are restricted to lie on a line 
(linear time, if $S$ and $T$ are presorted). 
\end{abstract}

\section{Introduction}
Consider two finite sets of points $S$ and $T$ with total 
cardinality $n$.  
The objective of the {\em assignment} problem is to establish a
correspondence between the points in $S$ and the points in $T$,
such that each point in $S$ corresponds to exactly one point in 
$T$, and each point in $T$ corresponds to at least one point 
in $S$. 
This correspondence is measured by a cost function $\delta$ that
assigns a cost $\delta(s, t)$ to each assigned pair $(s, t)$. 
The cost of an assignment is the sum of the costs of all assigned
pairs. The goal of the assignment problem is to find an assignment
of minimum cost.

The general assignment problem is also known as the 
{\em many-to-one assignment} problem. 
The {\em one-to-one} version of the assignment problem requires 
that each point in $S$ maps to exactly one point in $T$ and 
each point in $T$ gets mapped exactly one point in $S$.  
Throughout the paper, whenever we talk about the assignment
problem, we refer to the many-to-one version of the problem.

The simplest version of the assignment problem assumes
that the points in $S$ and $T$ lie on a line and the cost function 
is the $L_1$ metric.
In this setting, the one-to-one assignment problem 
has a simple $O(n \log n)$ time solution when $|S| = |T|$:
first sort the points in $O(n \log n)$ time, then map the $k^{th}$ point 
in $S$ to the $k^{th}$ point in $T$ in $O(n)$ time
\cite{bib:toussaintsimilarity}. 
However, the situation $|S| < |T|$ arises in many practical
applications. 
This situation was first addressed by Karp and Li~\cite{KL75},
who provided an $O(n \log n)$ time algorithm for the one-to-one 
assignment problem ($O(n)$ time, if $S$ and $T$ are given in 
sorted order). 
Simpler and equally efficient solutions have later been provided 
in ~\cite{ABKKS95, BY98, WPMK86}. 

Eiter and Mannila\cite{bib:eiter97distance} studied the assignment
problem in the context of measuring the distance between two
theories expressed in a logical language. They showed that for points
in arbitrary dimensions, this problem has a polynomial time solution. 
When restricted to points on a line, a minimum cost assignment 
can be used in measuring the similarity between musical 
rhythms. In this context, Toussaint~\cite{T03} proposed the use 
of the {\em directed swap distance} as a similarity measure.  
If the onsets of a rhythm are represented as points on a line 
separated by ``silence'' 
intervals, the directed swap distance between two rhythms with onset 
sets $S$ and $T$ is precisely the cost of an optimal 
assignment between $S$ and $T$, with underlying cost function $L_1$.

The assignment problem also appears in the shape of the 
{\em restriction scaffold assignment} problem in computational
biology~\cite{BKSS03}.
The goal here is to establish a correspondence between sparse 
experimental data and a restricted set of known structural 
building blocks. Ben-Dor et. al.~\cite{BKSS03} model the 
restriction scaffold assignment as an assignment problem
for points on a line, and provide an $O(n \log n)$ time algorithm to
solve this problem. However, as later shown by Colannino and 
Toussaint~\cite{CT05}, this algorithm fails to always produce
a minimum cost assignment. Thus, the best existing
solution to the assignment problem in one dimension
is the $O(n^2)$ algorithm presented in~\cite{CT05}.

In this paper, we show that the assignment problem 
with underlying cost function $L_1$ in one dimension 
can be solved in $O(n \log n)$ time 
($O(n)$ if the points in $S$ and $T$ are given in sorted order). 
Our algorithm is a simple extension of the $O(n \log n)$ time 
algorithm of Karp and
Li~\cite{KL75} for finding the minimum cost {\em one-to-one} 
assignment over $T$ and all subsets $S' \subset S$ of size $|T|$, 
assuming $|S| > |T|$. 
We present our algorithm in Section~\ref{sec:many-one}, 
after a few preliminary results (Section~\ref{sec:preliminaries}) 
and a close look at some properties of an optimal solution 
(Section~\ref{sec:properties}). 

\section{Background}
\label{sec:definition}
Let $S = \{s_0, s_1, s_2, \ldots\}$ and $T = \{t_0, t_1, t_2,
\ldots\}$ be two finite sets of points that lie on a horizontal line,
with $|S|+|T|=n$ and $|S| > |T|$.  For any $s \in S$ and $t \in T$,
the cost $\delta(s, t)$ of an assigned pair $(s, t)$ is the absolute value
of the difference between the $x$-coordinates of $s$ and $t$. To avoid
overloading the notation, we use the same symbol for a point and its
$x$-coordinate. Thus, $\delta(s, t) = |s - t|$. We assume that $s_i <
s_{i+1}, 0\leq i < |S|-1$ and $t_j < t_{j+1}, 0\leq j < |T|-1$.
 
An assignment $\A$ between $S$ and $T$
consists of pairs of points $(s, t)$ (henceforth {\em edges}), 
with $s \in S$ and $t \in T$, such that each point in $S$ belongs to 
exactly one edge in $\A$, and each point in $T$ belongs to at least 
one edge in $\A$. The cost of $\A$ is
\[ cost(\A) = \sum_{(s, t) \in \A} \delta(s, t) \] 
Our goal is to find an assignment $\A$ of minimum cost.
If two points in $S \cup T$ have the same $x$-coordinate, we can
slightly shift one of them to the left or right. If the minimum cost
assignment is unique and the change is sufficiently small, this change
will not affect the optimal assignment.  If there are several 
assignments with the same optimal cost, at least one of them 
will be the optimal solution of the new point set. So we may 
assume without loss of generality that all points in $S \cup T$ are 
distinct.

\subsection{Preliminaries}
\label{sec:preliminaries}
\begin{figure}[htbp]
\centering
\includegraphics[width=0.34\linewidth]{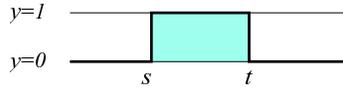}
\caption{Function $f_{s,t}$. Shaded area represents the 
cost $|s-t|$.}
\label{fig:intcost}
\end{figure}

For any $s \in S$ and $t \in T$, the value $|s - t|$ 
can be expressed in a different way as follows. 
Define a function $f_{s,t}$ to be $1$ in the interval 
between $s$ and $t$ and $0$ at any other 
point (see Figure~\ref{fig:intcost}). 
Then $|s - t| = \int_{-\infty}^{+\infty} f_{s,t}(x) dx$.

\noindent
The cost of an assignment $\A$ is therefore 
\[
  cost(\A) = \sum_{(s,t)\in \A} \int_{-\infty}^{+\infty} f_{s,t}(x) dx
           = \int_{-\infty}^{+\infty} \sum_{(s,t) \in \A} f_{s,t}(x) dx
\]
If we define
\[
  f_{\A}(x) = \sum_{(s,t) \in \A} f_{s,t}(x)  
\]
then the value $f_{\A}(a)$ is simply the number of 
edges in $\A$ pierced by the vertical line $x = a$, 
and the cost of $\A$ is

\begin{equation}
  cost(\A) = \int_{-\infty}^{+\infty} f_{\A}(x) dx
\label{eq:cost}
\end{equation}

\begin{figure}[htbp]
\centering
\includegraphics[width=0.6\linewidth]{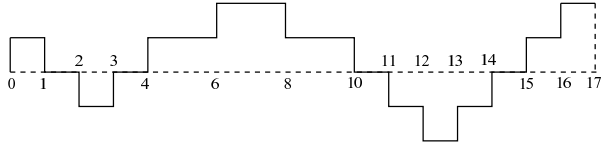}
\caption{Height function for sets 
$S = \{0, 3, 4, 6, 13, 14, 15, 16\}$ 
and $T = \{1, 2, 8, 10, 11, 12\}$.}
\label{fig:height}
\end{figure}

Our definition of $f_{\A}$ is similar in nature to the {\em height}
function $H:\mathbb{R}\rightarrow\mathbb{Z}$ introduced by Karp and Li
\cite{KL75}. Informally, they define $H(a)$ at each point $a$ as the
difference between the number of points in $S$ and the number of
points in $T$ restricted to the interval $(-\infty, a]$ (or
equivalently, to the left of the vertical line $x = a$).  Thus $H$
remains constant throughout each interval that does not contain a
point in $S \cup T$.  Figure~\ref{fig:height} shows the stair-shaped
curve of $H$ for a small example.
Note that {\em up} transitions in the curve correspond to 
points in $S$ and {\em down} transitions correspond
to points in $T$.  We refer to the value $H(x)$ as the {\em height} of
$x$. Note that $H(\infty) = |S|-|T|$.

\begin{lemma}
If $|S| = |T|$, then $\int_{-\infty}^{+\infty} |H(x)|~dx$ 
is the cost of the assignment that assigns the
$k^{th}$ largest element of $S$ to the $k^{th}$ 
largest element of $T$.
\label{lem:one}
\end{lemma}
\begin{proof}
Follows immediately from (\ref{eq:cost}) and the fact that,
for this particular assignment, $f_{\A}(x) = |H(x)|$ at each 
point $x$. 
\end{proof}

\begin{figure}[htbp]
\centering
\begin{tabular}{cc}
\includegraphics[width=0.6\linewidth]{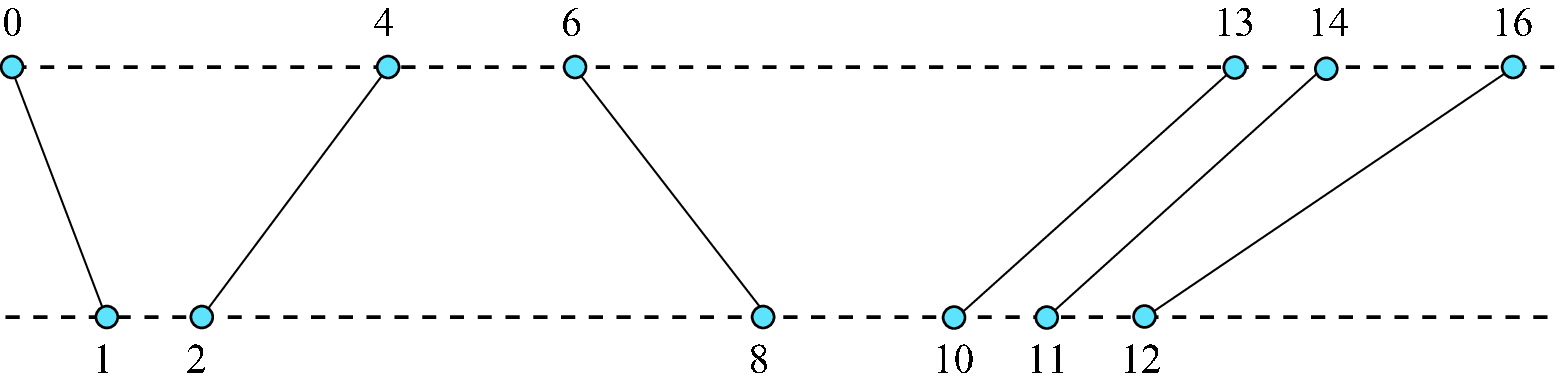} & 
\raisebox{7ex}{(a)} \\
 & \\
\includegraphics[width=0.6\linewidth]{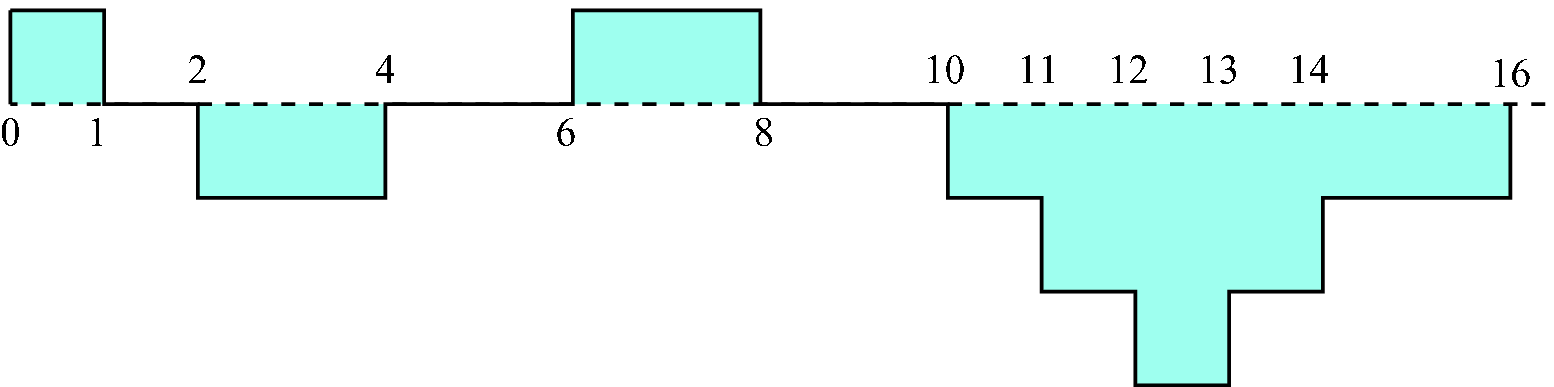} & 
\raisebox{7ex}{(b)}
\end{tabular}
\caption{
(a) One-to-one assignment for sets
$S = \{0, 4, 6, 13, 14, 16\}$ and 
$T = \{1, 2, 8, 10, 11, 12\}$
(b) Shaded area represents the cost of the assignment.}
\label{fig:equal.assign}
\end{figure}

\noindent
Figure~\ref{fig:equal.assign}a shows an assignment 
for two sets $S$ and $T$, with $|S| = |T|$. 
The cost of this assignment is equal
to the area shaded in Figure~\ref{fig:equal.assign}b,
which is precisely the value of the integral 
$\int_{-\infty}^{+\infty} |H(x)|~dx$. 

\section{Properties of a Minimum Cost Assignment}
\label{sec:properties}
Our algorithm for computing a minimum cost assignment $\A$
exploits several important properties of $\A$, which 
we discuss next.
A {\em crossing} is defined by a pair of 
edges $(a, d)$ and $(b, c)$ such that $a < b$ in $S$ 
and $c < d$ in $T$. 

\begin{figure}[htbp]
\centering
\begin{tabular}{c@{\hspace{0.06\linewidth}}
                c@{\hspace{0.06\linewidth}}c}
\includegraphics[width=0.22\linewidth]{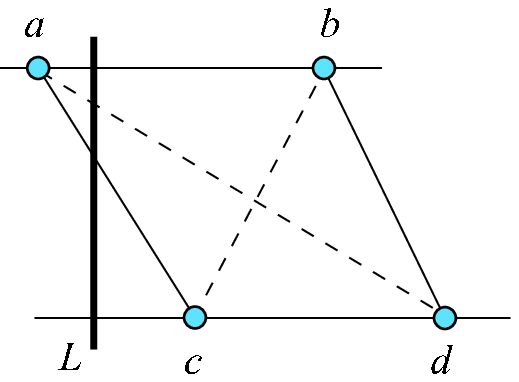} & 
\includegraphics[width=0.25\linewidth]{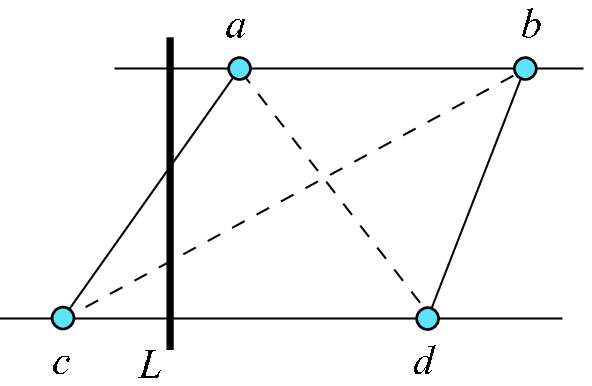} & 
\includegraphics[width=0.24\linewidth]{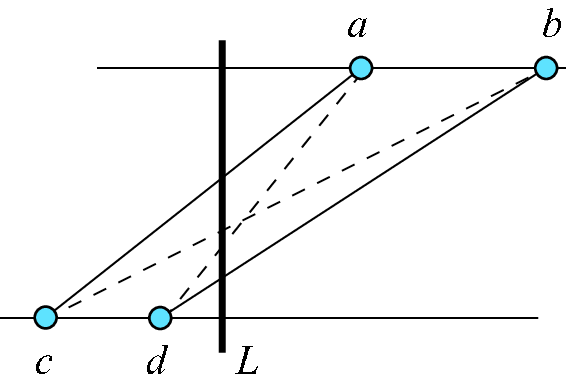} \\
(a) & (b) & (c)
\end{tabular}
\caption{
(a) Vertical line $L$ intersects $(a,c)$ and $(a,d)$ 
(b) $L$ intersects $(a, c)$ and $(b,c)$
(c) $L$ intersects 
$(a,c)$, $(b,d)$, $(a,d)$ and $(b,c)$.}
\label{fig:crossing}
\end{figure}

\begin{lemma}
There exists a minimum cost assignment with no crossings.
\label{lem:monotone}
\end{lemma}
\begin{proof}
Let $\A$ be a minimum cost assignment between $S$ and $T$ 
with a minimum number of crossings. If $\A$ has zero 
crossings, the proof is finished. Otherwise, pick 
two crossing edges $(a, d)$ and $(b, c)$ in $\A$, 
with $a < b$ in $S$ and $c < d$ in $T$.
We show that $\A' = \A \setminus \{(a, d), (b, c)\} 
\cup \{(a, c), (b, d)\}$ is an assignment with 
$cost(\A') \le cost(\A)$, a contradiction.
In particular, we show that 
$f_{\A'}(x) \le f_{\A}(x)$ at each point $x$; then
$cost(\A') \le cost(\A)$ follows immediately
from~(\ref{eq:cost}). 

First note that $f_{\A'}(x) \le f_{\A}(x)$ is true for 
any $x$ such that the vertical line $L$ at $x$ intersects 
neither of $(a,d)$ and $(b,c)$. 
Suppose now that $L$ intersects $(a,c)$. Then $L$ must 
also intersect either $(a, d)$ (see Figure~\ref{fig:crossing}a) 
or $(b, c)$ (see Figure~\ref{fig:crossing}b) or both
(see Figure~\ref{fig:crossing}c).
Similarly, if $L$ intersects $(b,d)$, then $L$ also intersects
at least one of $(a, d)$ and $(b, c)$. 
Furthermore, if $L$ intersects
both $(a, c)$ and $(b, d)$, then $L$ also intersects
both $(a, d)$ and $(b, c)$ (see Figure~\ref{fig:crossing}c).  
It follows that $f_{\A'}(x) \le f_{\A}(x)$.
\end{proof}

An assignment $\A$ can also be regarded as a function 
$\A : S \to T$ such that $\A(s) = t$ for each 
$(s, t) \in \A$. 
For any $t \in T$, let $\A^{-1}(t)$ denote the set of
elements $s \in S$ such that $\A(s) = t$. 
For each point $s \in S$, define the 
{\em nearest neighbor}  $N(s)$ to be point in 
$T$ closest to $s$, i.e, $|N(s) - s|\le |t - s|$
for any $t\in T$. In the case of a tie, $N(s)$ is
arbitrarily picked from among the two candidate
neighbors.

\begin{lemma}
Let $\A$ be optimal and let $t \in T$ be such that
$\A^{-1}(t)$ contains two or more elements. 
Then for each $s \in \A^{-1}(t)$, 
$t$ is a nearest neighbor of $s$. Furthermore, 
$T$ contains no points in between $s$ and $t$. 
\label{lem:neighbor}
\end{lemma}
\begin{proof}
Assume to the contrary that there is $s \in S$ with
$\A(s) = t, |\A^{-1}(t)| > 1,$ and $N(s) \neq t$. Refer to
Figure~\ref{fig:neighbor}. 
Define a new assignment $\A'$ with $\A'(s) = N(s)$ and 
$\A'(x) = \A(x)$ for $x \neq s$. Note that $\A'$ 
is also an assignment: 
$\A^{-1}(t)$ contains at least one point. 
Also $cost(\A') = cost(\A) - |s-t| + |s-N(s)|$ (see
Figures~\ref{fig:neighbor}a and~\ref{fig:neighbor}b). 
\begin{figure}[htbp]
\centering
\begin{tabular}{c@{\hspace{1in}}c}
\includegraphics[width=0.3\linewidth]{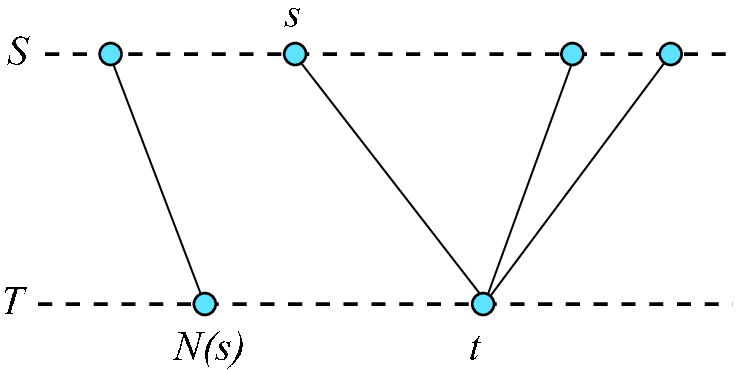} &
\includegraphics[width=0.3\linewidth]{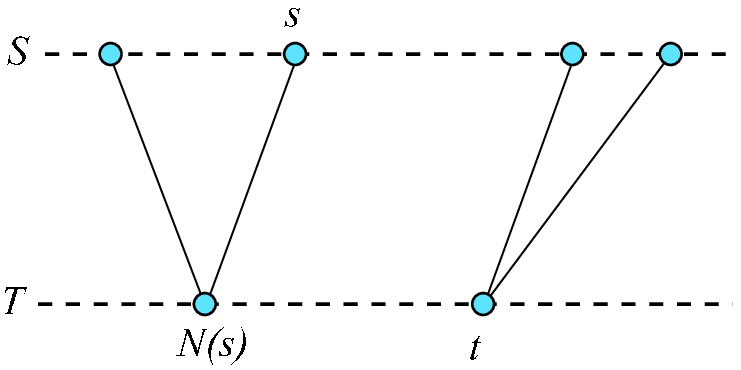} \\
(a) & (b)
\end{tabular}
\caption{
(a) Assignment $\A$ with $\A(s)\neq N(s)$
(b) Assignment $\A'$ with $\A'(s) = N(s)$}
\label{fig:neighbor}
\end{figure}
Since $|s-N(s)| < |s-t|$, it follows that 
$cost(\A') < cost(\A)$, 
contradicting the fact that $\A$ is of minimum cost.
Thus, $t$ is a nearest neighbor of $s$. 

The claim that $T$ contains no points in between 
$s$ and $t$ is immediate: if such a point $t_1\in T$ existed,
then $|s-t_1| < |s-t|$, 
contradicting the fact that $N(s) = t$.
\end{proof}

\medskip
\noindent
Observe that for any subset $R \subset S$ of size $|R| = |S|-|T|$,
there is a unique minimum cost assignment (with no crossings) 
from $S\setminus R$ to $T$. Let $\A_{S\setminus R}$ denote the 
edges of such an assignment, and define a new
assignment $\A_R: S\rightarrow T$ as follows:
\begin{equation}
\A_R(x) =
\begin{cases} N(x) & \text{if $x\in R$,}\\
y & \text{if $x\in S\setminus R$ and $(x,y)\in \A_{S\setminus R}$}
\end{cases}
\label{eq:ar}
\end{equation}

\vspace*{0.1in}
\noindent Lemma~\ref{lem:neighbor} implies that there always exists 
a subset $R$ such that $\A_R$ defines a minimum cost assignment 
from $S$ to $T$. Furthermore, $R$ satisfy a special height 
condition, stated in the lemma below. 

\begin{lemma}
There exists a subset $R\subset S$ with $|R|=|S|-|T|$ such that $\A_R$
defines a minimum cost assignment from $S$ to $T$, and the
$k^{th}$ smallest element of $R$ has height $k$.
\label{lem:height}
\end{lemma}

\begin{proof}
Let $\A:S\rightarrow T$ define a minimum cost assignment. We prove the 
existence of $\A_R$ by constructing a set $R \subset S$ 
with the properties stated in this lemma. Initially $R$ is empty.  
If $|\A^{-1}(t)| = 1$ for all $t\in T$, then $R$ is empty and 
the proof is finished.
Otherwise, we process points $t \in T$ for
which $\A^{-1}(t)$ has two or more elements. 
For each such point we consider two cases, 
as depicted in Figure~\ref{fig:lemma4cases}.  
If all points in $\A^{-1}(t)$ are less than $t$, then we 
add to $R$ all but the largest (rightmost) point in $\A^{-1}(t)$
(see Figure~\ref{fig:lemma4cases}a).
Otherwise, we add to $R$ all points in $\A^{-1}(t)$  
except for the smallest (leftmost) point greater than $t$
(see Figure~\ref{fig:lemma4cases}b).

\begin{figure}[htbp]
\centering
\includegraphics[width=0.7\linewidth]{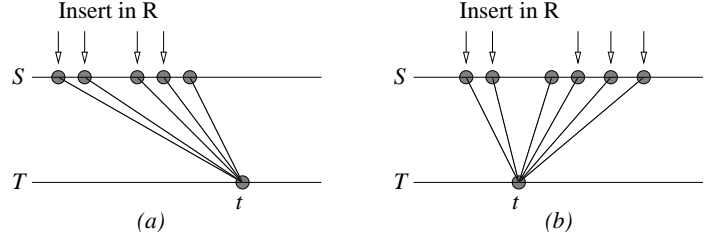} 
\caption{
(a) All points in $\A^{-1}(t)$ are less than $t$.
(b) Some points in $\A^{-1}(t)$ are greater than $t$.}
\label{fig:lemma4cases}
\end{figure}

We now define $A_R$ as in~(\ref{eq:ar}). Since $\A_R$
is identical to $\A$, $\A_R$ is a minimum
cost many-to-one assignment from $S$ to $T$.

It remains to show that 
the $k^{th}$ smallest element of $R$ has 
height $k$. To see this, first consider 
the smallest element of a nonempty set $\A^{-1}(t)\cap R$. Call this
element $r$ and suppose it 
is the $k^{th}$ smallest element of $R$. It follows then that (i)
$R$ contains  $k-1$ points less than $r$, and 
(ii) $T$ and $S \setminus R$ contain an equal number of 
elements less than $r$. This latter claim follows from 
Lemma~\ref{lem:neighbor}, which tells us that $T$ 
contains no elements in between $r$ and $t$, and the following
observation: the way in which we have selected $R$
ensures that if $t$ lies to the left of $r$ (i.e., $t<r$), the
assigned item for $t$ in $S/R$ lies to the 
left of $r$, and if $t$ lies to the right of $r$ ($t>r$), the assigned
item for $t$ in $S/R$ lies to the right of $r$.
These together imply that $H(r) = k$.  

We now show that the points in $\A^{-1}(t) \setminus \{r\}$ 
have height values $k+1, k+2, \ldots$, in order from 
smallest to largest.
By Lemma~\ref{lem:neighbor}, $T$ contains no points in between
$s$ and $t$, for each $s \in \A^{-1}(t)$. Then 
the points in $R \cap \A^{-1}(t)$ have incrementally 
increasing height values. It follows that the height of 
the $k^{th}$ smallest element of $R$ is $k$. 
\end{proof}

\medskip
\noindent
Let $H_R$ represent the height function restricted to
sets $S \setminus R$ and $T$. This means that for each $x$,
$H_R(x)$ is the 
difference between the number of points in $S \setminus R$ and the
number of points in $T$ restricted to the interval $(-\infty, x]$.

\begin{lemma}
The cost of 
assignment $\A_R$ is
\begin{equation} 
\sum_{r \in R} |r - N(r)| + \int_{-\infty}^{+\infty} |H_R(x)|dx 
\label{eq:removecost}
\end{equation}
\label{lem:optcost}
\end{lemma}
\begin{proof}
By Lemma~\ref{lem:one} we have that the contribution of 
$S \setminus R$ to the cost of $\A_R$ is 
$\int_{-\infty}^{+\infty} |H_R(x)|dx$. 
Since each point in $R$ maps to its nearest neighbor, the 
contribution of $R$ to the cost of $\A_R$ is 
$\sum_{r \in R} |r - N(r)|$. These together conclude the 
lemma.
\end{proof}

\begin{theorem}
Let $R \subset S$ be a subset of size $|R| = |S| - |T|$ with 
two properties:
\begin{enumerate}
\item[i.] The $k^{th}$ smallest element of $R$ has height $k$.
\item[ii.] $R$ minimizes the quantity from (\ref{eq:removecost}).
\end{enumerate}
Then $\A_R$ defines a minimum cost assignment from $S$ to
$T$.
\label{thm:properties}
\end{theorem}
\begin{proof}
By Lemma~\ref{lem:height}, we know that there exists a set $R$ that
satisfies {\em (i)}. By Lemma~\ref{lem:optcost}, 
$R$ satisfies {\em (ii)}. It follows that $\A_R$ is a minimum 
cost assignment from $S$ to $T$.
\end{proof}

\section{Computing a Minimum Cost Assignment}
\label{sec:many-one}
Theorem~\ref{thm:properties} gives an exact description of the set $R$
that yields a minimum cost assignment $\A_R$.  We now turn to the
problem of efficiently determining this set. With this goal in mind,
we introduce the following notation. 
For any point $x$ and any integer $k$, define the 
{\em relative height} of $x$ with respect to $k$ as 
\[ h^{k}(x) = \left\{ 
\begin{tabular}{ll}
$1$, & if $H(x) \ge k$ \\
$-1$, & if $H(x) < k$
\end{tabular}
\right.
\]
\noindent
Observe that when a point $s$ is removed from $S$, 
$H(x)$ decreases by 1 for all $x > s$. Suppose
that $H(s) = k$, and let $m$ be the largest point in $S \cup T$.
The removal of $s$ causes the area under the height
function between $s$ and $m$ to decrease by the quantity $\int_s^{m}
h^{k}(x) dx$. We use this observation to define the {\em profit} of
removing $s$ from $S$ and placing it in $R$ (recall that $\A_R$
assigns each item in $R$ to its nearest neighbor), as follows: 
\begin{equation}
P(s) = \int_s^{m} h^{k}(x) dx - |s - N(s)|
\label{eq:profdef}
\end{equation}

\begin{figure}[htbp]
\centering
\includegraphics[width=0.6\linewidth]{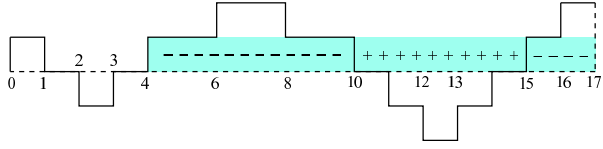}
\caption{A depiction of the integral $\int_s^{m}h^{k}(x) dx$ for $s =
4$.  The integral represents the effect of excluding $4$ from the
one-to-one assignment from $S$ to $T$.} 
\label{fig:relheight}
\end{figure}

The profit function quantifies the benefit of placing $s$ in $R$, the
goal being to minimize the cost of the assignment defined by $\A_R$. 
The integral term in~(\ref{eq:profdef}) represents the effect of 
excluding $s$ from the one-to-one assignment from $S\setminus R$ to $T$, 
as depicted in Figure \ref{fig:relheight}.  
The term $|s - N(s)|$ in~(\ref{eq:profdef}) represents the cost 
of assigning $s$ to its nearest neighbor.  
We minimize the cost of the assignment defined by $\A_R$ by
choosing items $s$ that maximize $P(s)$.
This is formalized in the following lemma.

\begin{lemma}
Let $R \subset S$ be a set with elements 
$r_1 < r_2 \ldots < r_{|S|-|T|}$ such that $H(r_k) = k$ 
and $r_k$ maximizes $P(s)$ among all points $s\in S$ of 
height $k$. Then $R$ minimizes
\[ \sum_{r \in R} |r - N(r)| + 
\int_{-\infty}^{+\infty} |H_R(x)|dx\]
\label{lem:cost.optimal}
\end{lemma}
\begin{proof}
Karp and Li~\cite{KL75} proved that any set $R$ of size $|S|-|T|$ whose 
$k^{th}$ smallest element has height $k$ satisfies the 
equality
\[\int_{-\infty}^{+\infty} |H_R(x)|dx = 
\int_{0}^m |H(x)|dx - 
\sum_{r \in R} \int_r^m h^{k}(x) dx\]
Summing up the cost contribution of $R$ to both sides of 
the equality yields 
\[ \sum_{r \in R} |r - N(r)| 
+ \int_{-\infty}^{+\infty} |H_R(x)|dx 
= 
\sum_{r \in R} |r - N(r)| 
+ \int_{0}^m |H(x)|dx 
- \sum_{r \in R} \int_r^m h^{k}(x) dx \]
This is equivalent to
\[\sum_{r \in R} |r - N(r)| 
+ \int_{-\infty}^{+\infty} |H_R(x)|dx 
= 
\int_{0}^m |H(x)|dx - \sum_{r \in R} P(r) \]
Since  $P(r_k)$ is maximized at each height $k$ and 
there is only one element in $R$ at each height, we have 
that $R$ maximizes $\sum_{r \in R} P(r)$, which in turn 
minimizes 
   \[\sum_{r \in R} |r - N(r)| 
     + \int_{-\infty}^{+\infty} |H_R(x)|dx\]
as required (refer to Lemma~\ref{lem:optcost}).
\end{proof}

The following algorithm uses the preceding lemma to determine
the optimal set $R$, and then compute the minimum cost 
assignment. 

\subsection{The Assignment Algorithm}
Initially $R$ is the empty set.
\begin{itemize}
\item[1.] Sort $S$ and $T$.
\item[2.] Calculate $H(x)$ for each $x \in S \cup T$. In between
consecutive points, $H$ is constant.
\item[3.] Calculate $P(s)$ for each $s \in S$.
\item[4.] For $k = 1, 2, \ldots |S|-|T|$
\begin{itemize}
\item[4.1] Find the leftmost point $r_k$ of height $k$ 
           that maximizes $P(r_k)$.
\item[4.2] Add $r_k$ to $R$.  
\end{itemize}
\item[5.] Return $\A_R$.
\end{itemize}

\begin{lemma}
The assignment algorithm computes a minimum cost assignment 
from $S$ to $T$. 
\end{lemma}
\begin{proof}
Let $r_k$ be the element of $R$ of height $k$ returned by the
algorithm. If we show that $r_1 < r_2 < \ldots < r_{|S|-|T|}$, then it
follows by Lemma~\ref{lem:cost.optimal} that $\A_R$ is a minimum cost
assignment. We prove below, by contradiction, that indeed $r_1
< r_2 < \ldots < r_{|S|-|T|}$.

Let $m$ be the largest point in $S$. 
Assume that there exists some $k (1\leq k\leq |S|-|T|-1)$ for which
the algorithm returns $r_k$ and $r_{k+1}$, with $r_k > r_{k+1}$.  Let
$s_k$ be the maximal element at height $k$ in $S \setminus R$ which is
less than $r_{k+1}$.  By continuity, such an $s_k$ must
exist. Similarly, let $s_{k+1}$ be the minimal element 
at height $k+1$ in $S \setminus R$ which is greater than $r_k$.  Such
an $s_{k+1}$ must exist since the height at $\infty$ is 
$H(\infty) = |S|-|T|$. Refer to Figure~\ref{fig:order}.
\begin{figure}[htbp]
\centering
\includegraphics[width=0.38\linewidth]{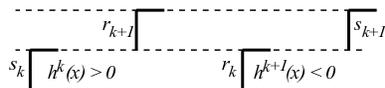} 
\caption{$s_k (s_{k+1})$ is the closest point at height $k (k+1)$ to
the left (right) of $r_{k+1} (r_k)$.}
\label{fig:order}
\end{figure}

Since
$H(r_{k+1}) = H(s_{k+1})$ and $r_{k+1} < s_{k+1}$, we have that
\[\int_{r_{k+1}}^{m}h^{k+1}(x)dx =
\int_{r_{k+1}}^{s_{k+1}}h^{k+1}(x)dx +
 \int_{s_{k+1}}^{m}h^{k+1}(x)dx\]
From this and equation~(\ref{eq:profdef}), we can derive the
following relation between the profit functions of $r_{k+1}$ and
$s_{k+1}$: 
\begin{equation}
   P(r_{k+1}) = P(s_{k+1}) + \int_{r_{k+1}}^{s_{k+1}} h^{k+1}(x) dx - 
                |r_{k+1} - N(r_{k+1})| + 
                |s_{k+1} - N(s_{k+1})|
\label{eq:profit1}
\end{equation}
Note that equality (\ref{eq:profit1}) is the result of breaking up the 
integral corresponding to $P(r_{k+1})$ into two parts, and taking 
into account the distance from each element to its nearest neighbor. 
Similarly, we can derive the following relation
between $P(r_k)$ and $P(s_k)$:
\begin{equation}
   P(s_k) =  P(r_k) + \int_{s_{k}}^{r_{k}} h^{k}(x) dx - 
             |s_{k} - N(s_{k})| + |r_{k} - N(r_{k})|
\label{eq:profit2}
\end{equation} 
The nearest neighbor of $s_k$ cannot be farther than
$N(r_{k+1})$. This translates into: 
\[ |s_k - N(s_k)| \leq |r_{k+1} - N(r_{k+1})| + 
|s_k - r_{k+1}| \] 
Also note that $h^k(x)$ is positive on the interval $(s_k, r_{k+1})$, 
which allows us to rewrite the previous equation as: 
\begin{equation}
|s_k - N(s_k)| \leq |r_{k+1} - N(r_{k+1})| + 
\int_{s_k}^{r_{k+1}}h^k(x) dx
\label{eq:profit3}
\end{equation}
Similar arguments lead to the following relationship between 
nearest neighbors of $r_k$ and $s_{k+1}$:
\begin{equation}
|r_k - N(r_k)| \geq |s_{k+1} - N(s_{k+1})| + 
\int_{r_k}^{s_{k+1}}h^{k+1}(x) dx
\label{eq:profit4}
\end{equation}
Finally, on the interval $(r_{k+1}, r_k)$ note that 
\begin{equation}
\int_{r_{k+1}}^{r_{k}}h^{k+1}(x) dx \leq 
\int_{r_{k+1}}^{r_{k}}h^{k}(x) dx 
\label{eq:profit5}
\end{equation}
Let $M_k = |s_{k} - N(s_{k})| - |r_{k} - N(r_{k})|$. 
Simple arithmetic that involves inequalities (\ref{eq:profit3}), 
(\ref{eq:profit4}) and (\ref{eq:profit5}) yields
\[\int_{s_{k}}^{r_{k}} h^{k}(x) dx - M_k 
\geq \int_{r_{k+1}}^{s_{k+1}} h^{k+1}(x) dx + M_{k+1} \]
This along with (\ref{eq:profit1}) and (\ref{eq:profit2}) 
implies that 
\[ P(s_k) - P(r_k) \geq P(r_{k+1}) - P(s_{k+1})\]

Since $r_{k+1}$ was picked by the assignment algorithm, we have that
$P(r_{k+1})\geq P(s_{k+1})$. This implies that $P(s_k)\geq P(r_k)$,
but since $s_k$ lies to the left of $r_k$, the assignment algorithm would
have picked $s_k$ instead of $r_k$, a contradiction.
\end{proof}

\subsection{Complexity Analysis}
Sorting in step 1 takes $O(n \log n)$ time.  All other steps run in
$O(n)$ time.  The only steps where this is not obvious are steps 2 and
3 that involve computing $H(x)$ and $P(x)$ respectively.  
$H(x)$ can be computed
for all $s \in S$ by conducting a sweep of the sorted points in $S
\cup T$, adding one when we encounter an element of $S$ and
subtracting one when we encounter an element of $T$.

Since all nearest neighbors of the elements of $S$ can easily be computed in 
linear time, to show that we can compute the profit function for all 
elements of $S$ in linear time we concern ourselves only with computing 
the integral of relative height function $h^k$.  This 
integral can be computed in linear time for all points in $S$ at 
height $k$ in a sweep from right to left.  
For the rightmost element $s_r$ of $S$ at height $k$ 
$\int_{s_r}^{m}h^{k}(x)dx = |s_r - m|$, where $m$ is the largest 
point in $S$.  Suppose that we know $\int_{s}^{m}h^{k}(x)dx$ for some
item $s$ at height $k$. 
Let $s' < s$ be the largest element in $S$ 
also at height $k$, and let $t < s$ be the largest element 
in $T$ at height $k$.  Note that by continuity, $t$ exists and
must be greater than  $s'$.  Also note that $h^k(x)$ is positive for
all $s'\leq x\leq t$, and $h^k(x)$ is negative for all $t<x<s$.
Thus we can derive the following equation:
\begin{equation}
\int_{s'}^{m} h^k(x)dx = \int_{s}^{m} h^k(x)dx + |s' - t| - |t - s|
\end{equation}
This value can be computed in constant time for each $s' \in S$.  
Thus we can compute $P(s)$ for all $s \in S$ in linear time.

It follows that the assignment algorithm runs in $O(n \log n)$
time. Furthermore, if $S$ and $T$ are given in sorted order, 
the assignment algorithm runs in $O(n)$ time. 

\section{Conclusion}
We have shown that the one-to-one assignment algorithm in ~\cite{KL75}
can be extended to produce a minimum cost many-to-one assignment. The
algorithm runs in $O(n \log n)$ time, if the input points are given in 
arbitrary order, and in $O(n)$ time, if the input points are presorted. 
To our knowledge, this is the first solution to the assignment problem 
that achieves this time complexity.

\end{document}